%% file: samplepaper.tex
\documentclass[runningheads]{llncs}
\usepackage[T1]{fontenc}
\usepackage{graphicx}
\usepackage[hidelinks]{hyperref}
\usepackage{tikz}
\usetikzlibrary{positioning, shapes.geometric, arrows.meta}
\usepackage{booktabs}
\usepackage{float}
\usepackage{placeins}
\usepackage{pgfplots}
\pgfplotsset{compat=1.18}
\usepackage{listings}
\usepackage{orcidlink}
\input{config}
\usepackage{cleveref}

\begin{document}
\title{Automated LLM-Based Accessibility Remediation: From Conventional Websites to Angular Single-Page Applications\thanks{This research has been supported by MICIU/AEI/10.13039/501100011033 under projects PLEC2023-010266 (SOFIA), PID2022-142964OA-I00 (EGSVAI) and RED2022-134647-T (AI4Software). It has also been partially funded by European Regional Development Funds (ERDF) and the University of Malaga.}}
\titlerunning{Automated Accessibility Remediation for Web and Angular SPAs}
%
\author{Carla Fernández-Navarro\orcidlink{0009-0004-4720-0453} \and
Francisco Chicano\orcidlink{0000-0003-1259-2990}}
\authorrunning{Fernández-Navarro and Chicano}
%
\institute{ITIS Software, University of Malaga\\
\email{\{carlafdez12,chicano\}@uma.es}}
\maketitle              
\begin{abstract}
Web accessibility remains an unresolved issue for a large part of the web content. There are many tools to detect errors automatically, but fixing those issues is still mostly a manual, slow, and costly process in which it is easy for developers to overlook specific details. The situation becomes even more complex with modern Single-Page Applications (SPAs), whose dynamic nature makes traditional static analysis approaches inadequate. This work proposes a system that aims to address this challenge by using Large Language Models (LLMs) to automate accessibility fixes. The proposal presents a modular workflow applicable to both static websites and complex Angular projects. The framework actively implements corrections within the DOM of static web pages or the source code of SPAs. The system was tested on 12 static websites and 6 open-source Angular projects, fixing 80\% of the accessibility issues on public websites and 86\% of the issues on Angular applications. Our proposal also generates meaningful visual descriptions for images while preserving the application's design and stability. This work contributes to ensuring that accessibility stops being a technical debt deferred to the future and becomes a natural part of everyday development workflows.

\keywords{Web accesibility  \and Automatic remediation \and Generative AI \and Large Language Models.}
\end{abstract}
\section{Introduction}
Digital exclusion remains a pervasive issue in the modern web. Despite legal mandates like the European Accessibility Act\footnote{\scriptsize\url{https://commission.europa.eu/strategy-and-policy/policies/justice-and-fundamental-rights/disability/european-accessibility-act-eaa_en} (Last Access: January 2026)}, millions of users are blocked by interfaces that fail basic compliance. The aim is to ensure equal access to digital materials and services for all users, regardless of their physical or cognitive capabilities. Millions of people with visual, auditory, physical, or mental impairments are effectively excluded by these barriers, making it impossible for them to fully engage with the digital world. Axe-core\footnote{\url{https://www.deque.com/axe/axe-core/} (Last Access: February 2026)} and WAVE\footnote{\url{https://wave.webaim.org/} (Last Access: February 2026)} are two automated tools that can help identify online accessibility concerns. These tools have potential, but their effectiveness is limited because they frequently remain ``fail to capture semantic or context-dependent violations'' \cite{data10090149} related to user experience. Consequently, many websites still have significant accessibility barriers that prevent users from accessing and utilizing them.

Modern Single-Page Applications (SPAs) add another layer of complexity to this challenge. Frameworks like Angular create dynamic, client-side rendered content, which introduces complex accessibility issues (e.g., focus management, route announcements) that traditional static analysis tools struggle to evaluate effectively \cite{10.1145/3594806.3596542}. Moreover, while error detection is well-supported, remediation of these issues is often the major bottleneck. Manual accessibility remediation requires significant human effort, can be time-consuming and resource-intensive, and may not result in fully accessible websites \cite{10.1145/3594806.3596542}.

Although some studies have explored using Large Language Models (LLMs) such as OpenAI for web accessibility remediation as a case study \cite{huang2024accesspromptengineeringautomated}, a fully automated end-to-end tool that integrates detection and correction, especially for SPAs, remains largely unexplored. In this paper, we present an automatic tool that utilizes LLMs for the automatic remediation of web accessibility. We explored the effectiveness of this model as an automated tool by applying it to both traditional static websites and SPAs built with Angular, focusing on the Web Content Accessibility Guidelines 2.2 conformance Levels A and AA.

Accessibility issues of the selected websites were detected using the Axe-core library injected via Selenium \footnote{\url{https://www.selenium.dev/} (Last Access: January 2026)}. Our decision to use Axe-core was influenced by its reliability, ease of integration, and widespread industry adoption. Once the areas of non-compliance were identified, the selected websites were subjected to remediation using our tool, which leverages the OpenAI API. The system provides the model with the necessary technical context, including the identified violations and the relevant code segments—to generate corrected HTML and context, aware image descriptions. The effectiveness and accuracy of the obtained outcomes are subsequently evaluated by re-analyzing the corrected web page. The resulting findings could have important implications for web developers and designers. By adopting this automated remediation approach, stakeholders can potentially save time and resources while increasing the accessibility of both their traditional and modern SPA websites.

The remainder of this paper is organized as follows. Section~\ref{sec:literature-review} presents a review of the literature on web accessibility, automated testing tools, and specific challenges posed by SPAs. Section~\ref{sec:methodology} explores the architecture of our proposed tool. Section~\ref{sec:evaluation} describes the implementation details and Section~\ref{sec:results} presents the obtained results from our validation on websites and SPAs. Finally, Section~\ref{sec:conclusions} concludes the paper and highlights future research directions.

\section{Literature review}
\label{sec:literature-review}
Our research is situated at the intersection of three rapidly evolving fields: the limitations of traditional accessibility evaluation, the challenges posed by modern web applications, and the emerging use of Large Language Models (LLMs) for code remediation. In the next paragraphs, we will cover the recent advances in these three fields.
 
Despite the widespread adoption of the Web Content Accessibility Guidelines (WCAG), their manual implementation is expensive and error-prone. Automated tools such as WAVE are well-established for their ability to deliver detailed and reliable assessments of a website's accessibility \cite{10.1145/3663547.3746360}. Nevertheless, these tools are essentially diagnostic; they find infractions, but the developer has the entire responsibility for fixing them. The ``detect-then-correct'' process, which is frequently done by hand, continues to be a major obstacle to creating a more accessible online experience.

This reliance on manual correction has led to a clear need for automated remediation. Early research, such as the case study by Achraf Othman et al. \cite{10.1145/3594806.3596542}, demonstrated that general-purpose LLMs, including ChatGPT, could be used to address problems found by tools such as WAVE \cite{10.1145/3706598.3713335}. This work established a baseline, showing that LLMs could understand the intent of an accessibility violation and suggest a workable code remedy, moving beyond simple static~analysis~\cite{10.1145/3594806.3596542}.

However, the scope of these solutions must be expanded to address accessibility challenges in dynamic Single-Page Applications, where detection and remediation of accessibility issues are considerably more complex. The dynamic, client-side rendering present in SPAs was beyond the capabilities of traditional static analyzers, which were made for static, server-rendered material. Consequently, these tools are limited in their ability to evaluate stateful environments generated by dynamic JavaScript and AJAX \cite{huang2024accesspromptengineeringautomated}. Addressing accessibility in SPAs requires techniques capable of exploring application states that emerge only at runtime \cite{data10090149}.
Runtime behaviors present a significant challenge in SPAs. For example, in Angular applications, client-side routing does not trigger a full-page reload. Because of this, screen reader users must carefully manage their focus to avoid becoming disoriented \cite{unknown}. Such runtime-dependent accessibility requirements cannot be reliably captured through static analysis of HTML files alone. Therefore, dynamic testing and remediation strategies are essential to ensure that SPAs meet accessibility standards \cite{unknown2}.

In parallel with these structural challenges, the rise of LLMs for automated remediation has recently become a dedicated field of research. The ``ACCESS'' initiative by Huang et al. \cite{huang2024accesspromptengineeringautomated} moved beyond simple case studies to focus on \textit{prompt engineering} as a systematic method for automated web accessibility corrections. This signaled a change towards creating reproducible and reliable methods for using LLMs in this domain.

Other strategies focus on integrating accessibility during the development process. For instance, the ``CodeA11y'' framework seeks to make GitHub Copilot and other AI coding aids useful for accessible web development \cite{10.1145/3706598.3713335}. This technique reinforces best practices for developers by attempting to stop accessibility problems from being written in the first place by offering real-time feedback.

The current state of the art is moving towards comprehensive remediation tools. The ``AccessGuru'' project \cite{10.1145/3663547.3746360} combines existing testing libraries with LLMs to not only detect but also correct violations, creating a novel taxonomy of violation types to improve correction accuracy. Supporting this, Andruccioli et al. \cite{data10090149} developed a specialized dataset to benchmark LLMs on this task, calling it a ``paradigm shift'' from conventional static analyzers. While their work acknowledges modern frameworks such as React and Angular, it serves as a static evaluation resource rather than a real-time intervention tool.

Despite these advancements, we identified a significant research gap in this field. While LLMs show promise in controlled environments, applied research lacks robust validation on ``live'' frameworks like Angular, where the DOM is volatile. On the one hand, studies show that SPAs pose special, dynamic difficulties, such as focus management, which are significant obstacles for people with visual and motor impairments. Critically, the dynamic runtime problems that most impact users with impairments on SPAs are not addressed by current LLM solutions, which are trained on static HTML.

Our work directly addresses this gap by presenting a versatile end-to-end remediation framework designed for both traditional web environments and dynamic Single-Page Applications (SPAs). Unlike previous approaches often limited to static analysis, our tool operates on the live, dynamic DOM of Angular applications while remaining highly effective for conventional static websites. By combining Selenium-based interaction with an LLM-powered correction engine, the framework bridges the gap between offline auditing and the real-world accessible web, proving that automated remediation can handle the structural simplicity of static pages and the complex runtime challenges of modern frameworks with equal robustness.

\section{Methodology and Tool Design}
\label{sec:methodology}

To address the gap between automated detection and remediation, we designed a modular software tool capable of analyzing web interfaces and automatically generating corrected HTML code. The system identifies accessibility violations based on WCAG 2.2 Level A. Additionally, adhering to the guidelines of the Spanish Web Accessibility Observatory\footnote{\url{https://administracionelectronica.gob.es/pae_Home/pae_Estrategias/pae_Accesibilidad/pae_Observatorio_de_Accesibilidad.html} (Last Access: January 2026)}, the remediation for Level AA is explicitly restricted to five specific criteria: Reflow, Multiple Ways, Device Independence, Language of Parts, and Consistent Navigation. The system was developed in Python using a component-based architecture to ensure scalability and maintainability.

\begin{figure}[htbp]
    \centering
    \includegraphics[width=1\textwidth]{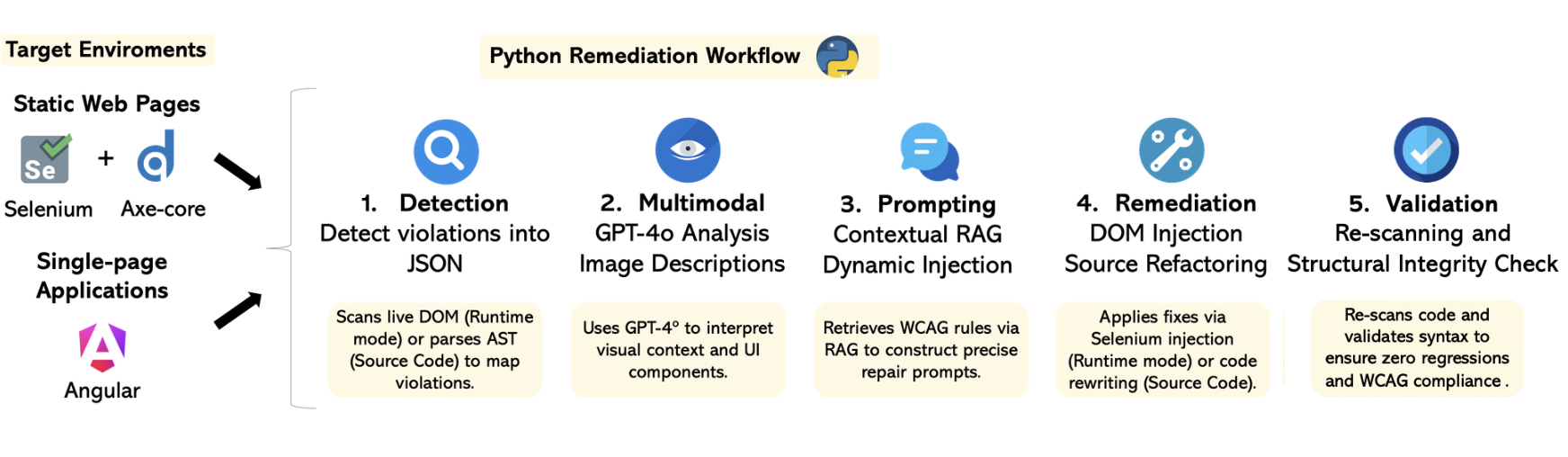}
    \caption{Pipeline of the repair tool}
    \label{fig:pipeline}
\end{figure}

Following the remediation framework paradigm explored in recent studies~\cite{huang2024accesspromptengineeringautomated}, our system operates as an end-to-end pipeline composed of four distinct phases: Detection and Discovery, Multimodal Visual Analysis and Context, Contextual Prompt Construction, and Remediation Execution (see Figure~\ref{fig:pipeline}).

A key contribution of this project is the implementation of a {modular system} that adapts these phases to two distinct target environments: {Static Web Pages} for public websites using Selenium, and {Single-Page Applications} for local Angular projects. The following subsections provide a detailed explanation of the four core stages of our remediation workflow: Detection, Multimodal Analysis, Prompting, and Remediation.

\subsection{Detection and Discovery} The first phase involves identifying non-compliant elements or components, depending on the target environment:
\begin{itemize}
    \item For static webs, the system utilizes a headless browser controlled via \textit{Selenium} to render the full Document Object Model (DOM). This ensures that elements generated via JavaScript are present in the structure being audited. The detection engine injects \textit{Axe-core} into the live DOM to scan for WCAG~2.2 Levels A and AA violations, extracting the specific node location and violation data.
    \item  For Angular projects, the tool parses the \texttt{angular.json} configuration file to identify the project's internal directory structure. It performs a recursive traversal to build a dependency graph of the application, identifying the triad of files associated with each component: the HTML template (\texttt{.html}), the TypeScript code (\texttt{.ts}), and the styles (\texttt{*.css}). The detection phase in this mode is conducted through a static analysis engine that identifies non-compliant patterns within the component templates, cross-referencing them with WCAG 2.2 standards and Axe-core's rule definitions before passing the context to the LLM for remediation.
\end{itemize}

\subsection{Multimodal Visual Analysis and Context}

To resolve WCAG criterion 1.1.1 (Non-text Content), the system incorporates a multimodal analysis phase. It identifies images lacking descriptions, downloads them to a local cache, and processes them using the GPT-4o multimodal model. As shown in Fig.~\ref{lst:prompt_vision}, a specific multimodal prompt is used to request a concise and helpful description for a screen reader. These descriptions are cached and  injected into the remediation context to be placed within the \texttt{alt} attributes.

Furthermore, during the Multimodal Analysis phase and prior to generating any remediation, the tool captures full-page screenshots across multiple viewports (Mobile, Tablet, and Desktop). These visual artifacts are injected into the context of the Holistic Remediation Prompt (see Appendix A, Listing~\ref{lst:prompt_angular} and \ref{lst:prompt_vision}). This allows the model to analyze layout-dependent elements and verify color contrast issues in their rendered state, ensuring that the suggested fixes better interpret the visual hierarchy—such as element overlapping or alignment—that a purely code-based analysis might miss.

\subsection{Contextual Prompt Construction} 

Once violations of elements are identified, the system constructs the remediation query for the LLM. Unlike approaches that rely on generic queries, our tool uses pre-defined, specialized prompt templates that are dynamically populated with the specific context.

The prompt template is designed to minimize ``hallucinations'' and maximize technical accuracy. As illustrated in Fig.~\ref{lst:prompt_structure}, each query is composed of three parts:
\begin{itemize} 
\item \textbf{System Instruction:} A high-level directive to instruct the LLM to act as a WCAG 2.2 expert. This establishes the model's persona and remediation constraints. 
\item \textbf{Dynamic Context Injection:} The engine programmatically retrieves and inserts the relevant code snippets, Axe-core violation data, or Angular logic. This process is detailed in the \textbf{Prompting} phase of our workflow. 
\item \textbf{Output Constraints:} Instructs the LLM to use strict formatting rules, such as the use of \verb|<<<TEMPLATE>>>| delimiters, to ensure the response can be parsed and applied automatically to the source files or the DOM. 
\end{itemize}

\subsection{Remediation Execution} 

The final phase applies the corrections, managing the lifecycle of the target environment. The system distinguishes between Static Web Pages, where fixes are applied as a temporary patch within the browser's session, and Single-page Applications, where the tool acts as an automated developer assistant to permanently modify the project's files.
We next detail both alternatives.

In the case of HTML static web pages, we implement Dynamic DOM Stabilization. Using \textit{Selenium's WebDriverWait}, the tool monitors network and DOM activity to ensure the page is fully rendered. Once fixes are generated our tool uses the prompt in Figure~\ref{lst:prompt_contrast} for contrast issues and the one in Figure~\ref{lst:prompt_general} for general issues. Then, \textit{BeautifulSoup} is used to surgically replace the defective nodes in the live DOM with the accessible versions, generating an HTML artifact. To maintain responsive integrity during this process, the system performs a final merge using the prompt in Figure~\ref{lst:prompt_responsive}.

For Angular SPA projects, the tool implements remediation at the development level. Unlike the static mode which modifies the final rendered output, this approach targets the raw component files. The LLM returns the code organized into structured segments, using specific delimiters such as \verb|<<<TEMPLATE>>>| and \verb|<<<TYPESCRIPT>>>| to isolate the HTML and code blocks, as specified in Figures~\ref{lst:prompt_template} and~\ref{lst:prompt_angular}. These markers allow the system to precisely parse the AI's response, ensuring that only the relevant code is extracted and injected into the corresponding file without affecting the rest of the component's structure. The system validates these segments and performs a differential write-back to the actual source files. This ensures that accessibility improvements are integrated into the version control system and fully preserved throughout the Angular compilation process and future build cycles.

\subsection{Validation of Remediation Accuracy} 

\begin{figure}[!ht]
    \centering
    \includegraphics[width=0.9\textwidth]{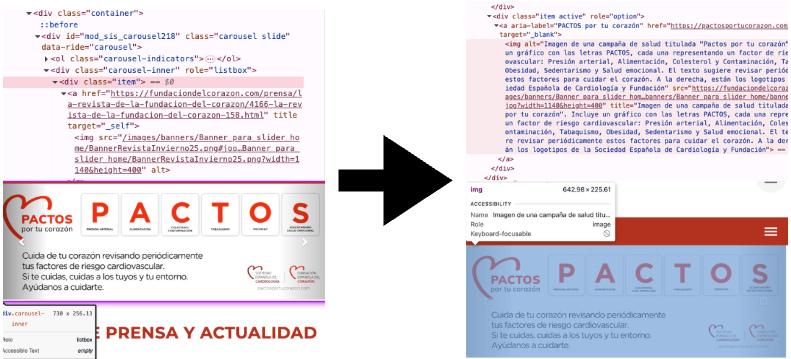}
    \caption{Multi-stage Validation and Verification Pipeline for AI-Generated Fixes. We can see our tool provides the content of the \texttt{alt} attribute of an image.}
    \label{fig:example}
\end{figure}

To ensure that each generated fix is both syntactically valid and compliant with the intended WCAG criteria, the system incorporates a multi-stage validation process. After the remediation is applied, the corrected HTML or component template is re-evaluated using Axe-core to verify that the original violations have been resolved and that no additional issues have been introduced. This transition is illustrated in Fig.~\ref{fig:example}, which shows a comparison between the original non-compliant HTML and the final validated output. As observed in the figure, the tool successfully transforms a complex image-based carousel item lacking descriptive context into an accessible element. The remediation includes the injection of a precise \texttt{aria-label} for the link and a comprehensive \texttt{alt} attribute for the image, which was generated through the multimodal analysis of the visual content. For Angular projects, the system utilizes BeautifulSoup to validate the structural integrity of the HTML segments and RegEx-based parsers to verify the syntax of TypeScript blocks before the write-back occurs. Additionally, the tool monitors the Angular Compiler (ng build/serve) output in the development environment to catch any potential build-time errors. If any stage of this validation pipeline fails—whether due to persistent accessibility violations, structural malformation, or compilation errors—the proposed changes are automatically discarded, and the specific remediation attempt is logged as a failure to prevent the introduction of regressions into the production or source code. This combination of automated re-scanning and structural validation ensures that the LLM-produced corrections are safe, consistent, and maintainable.

\FloatBarrier

\FloatBarrier

\section{Experimental Evaluation}
\label{sec:evaluation}

The evaluation protocol followed in this work is based on methodologies of recent studies on automated remediation and 
%
%
is designed to answer three fundamental research questions that address remediation efficacy and source code integrity of the proposed tool:
\begin{itemize}
\item \textbf{RQ1 (static web pages remediation efficacy):} To what extent can the tool automatically identify and fix WCAG 2.2 Level A and AA violations in complex, real-world static web pages without human intervention? To measure the remediation efficacy, we will calculate the percentage of accessibility violations successfully fixed and validated by Axe-core after the tool's execution, comparing this to the initial total of detected violations.
\item \textbf{RQ2 (Angular SPAs remediation efficacy):} Is the tool capable of modifying Angular SPA source code to improve accessibility while preserving the application's compilation validity? This is evaluated through Build-Success Verification. The tool initiates the Angular Ahead-of-Time (AOT) compiler after completing the differential write-back. We need the project to maintain a clean build following remediation in order to validate the results. This phase confirms that the tool hasn't affected the stability of the program and that the changes are syntactically sound.
\item \textbf{RQ3 (semantic preservation):} Do AI-driven fixes prevent ``destructive fixes'' and maintain the interface's original functionality and visual design?  We verify this by performing a manual side-by-side comparison of the application before and after the fixes. The goal is to ensure that accessibility enhancements do not interfere with the existing CSS styling or break the component's logic (e.g., event listeners or data binding). If a modification shifts the visual structure or disables interactive elements, it is marked as a failure. This ensures that the original UI remains intact throughout the process.
\end{itemize}

To answer these questions, we collected two different datasets (presented in the next subsection), ensuring a diverse representation of web technologies and architectural patterns, and we used the evaluation metrics presented in Subsection~\ref{subsec:metrics}. For the sake of reproducibility, both the complete datasets and the source code of our tool are publicly available on the Zenodo repository\cite{replication-package}

\subsection{Datasets}

To analyze remediation efficacy for static web pages, we select a diverse set of 12 public websites, called Dataset A, presented in Table~\ref{tab:dataset_a_urls}, classified into two strategic sectors to ensure the technological and semantic variety:
\begin{itemize} 
\item \textbf{Healthcare sector:} This group includes regional health service portals, specialized hospital websites, and a heart disease prevention platform. These sites were selected due to the critical nature of ensuring access for patients with disabilities and the rigorous WCAG compliance mandates governing the public healthcare sector. 
\item \textbf{Academic and research sector:} This subset consists of research group portals, academic blogs, and international scientific conference websites. These sites are essential for evaluating the tool's robustness against ``legacy'' HTML structures, domain-specific technical terminology, and complex data visualizations that frequently present significant accessibility barriers. 
\end{itemize}

\begin{table}[!ht]
\centering
\caption{Selection of public websites for Dataset A categorized by sector}
\label{tab:dataset_a_urls}
\scalebox{0.85}{
\begin{tabular}{@{}cllc@{}}
\toprule
\textbf{ID} & \textbf{URL / Website} & \textbf{Sector} \\ \midrule
1 & \url{oplink.lcc.uma.es} & Academic \& Research \\
2 & \url{precog.lcc.uma.es} & Academic \& Research \\
3 & \url{6city.lcc.uma.es} & Academic \& Research \\
4 & \url{smart-ct2017.lcc.uma.es} & Academic \& Research \\
5 & \url{smart-ct2016.lcc.uma.es} & Academic \& Research \\
6 & \url{blogs.ugr.es/musicaygenero/} & Academic \& Research \\ \midrule
7 & \url{hospitalpuertadelmar.com} & Healthcare \\
8 & \url{quironsalud.com/valencia} & Healthcare \\
9 & \url{hospitaldelmar.cat/es/} & Healthcare \\
10 & \url{fundaciondelcorazon.com} & Healthcare \\
11 & \url{sanidad.castillalamancha.es} & Healthcare \\
12 & \url{vithas.es/centro/vithas-hospital-granada/} & Healthcare \\ \bottomrule
\end{tabular}
}
\end{table}

To analyze single-page applications, we select a diverse set of 6 open-source Angular projects, called Dataset B, hosted in GitHub. The projects, presented in Table~\ref{tab:dataset_b_projects}, were selected due to their high number of forks and starts in GitHub (see columns 3 and 4 in the table).

\begin{table}[!ht]
\centering
\caption{Selection of open-source Angular projects for Dataset B}
\label{tab:dataset_b_projects}
\scalebox{0.85}{
\begin{tabular}{@{}clrr@{}}
\toprule
\textbf{ID} & \textbf{Project Repository (GitHub)} & \textbf{Forks} & \textbf{Stars} \\ \midrule
1 & \url{tomalaforge/angular-challenges} & 2700 & 1400 \\
2 & \url{angular-university/angular-ssr-course} & 153 & 196 \\
3 & \url{angular-university/angular-course} & 667 & 424 \\
4 & \url{trungvose/jira-clone-angular} & 587 & 2300 \\
5 & \url{realworld-apps/angular-realworld-example-app} & 3400 & 5600 \\
6 & \url{mauriciovigolo/keycloak-angular} & 315 & 846 \\ \bottomrule
\end{tabular}
}
\end{table}

\subsection{Evaluation Metrics}
\label{subsec:metrics}
To quantify the answers to the proposed research questions, we defined the following metrics:

\begin{itemize}
    \item \textbf{Remediation Rate (RR):} It measures the percentage of resolved accessibility violations, calculated as:
    \begin{equation}
        RR = \left( \frac{V_{initial} - V_{final}}{V_{initial}} \right) \times 100,
    \end{equation}
    where $V_{initial}$ is the violation count from the Axe-core baseline scan, and $V_{final}$ is the count after the automated remediation process. This metric is used to answer RQ1 and RQ2. 

    \item \textbf{Build Integrity (BI):} A binary metric (Pass/Fail) that validates if the Angular project compiles successfully using the Angular CLI (\verb|ng build|) after the source code modifications. This metric is used to answer RQ2. 

    \item \textbf{Semantic Functional Verification (SFV):} A qualitative assessment based on manual sampling to verify that critical UI elements (buttons, navigation menus) retain their interactive functionality. This metric is used for RQ3. 
\end{itemize}

\section{Results}
\label{sec:results}

This section details the results obtained by the tool across the two datasets. We use these results to answer the three research questions, presented in different subsections.

\subsection{RQ1: Static Web Pages Remediation Efficacy}
We applied our tool to Dataset A (the 12 public websites in Table~\ref{tab:dataset_a_urls}). Table \ref{tab:individual_results} presents the detailed breakdown of violations detected and fixed for each website.
The results demonstrate a high degree of efficacy for modern implementations. In 8 out of the 12 cases, the tool achieved a 100\% remediation rate (fixing all detected Level A and AA issues).
However, the results also reveal significant outliers linked to the website's age and technology:
\begin{itemize}
\item \textbf{Case 11 (sanidad.castillalamancha.es):} This site presented a lower remediation rate of 23.81\%. Manual inspection revealed that the site relies on complex legacy structures and non-standard containers, a practice that prevents the LLM from inferring the correct semantic context for ARIA injection.
\item \textbf{Cases 7, 8 \& 12 (hospitals):} The healthcare portals (Hospital Puerta del Mar, Hospital Quirón de Valencia, and Hospital de Mar) showed varied violation counts. The tool fixed between 45\% and 47\% of errors in these cases, with the remaining issues confined to complex third-party widgets and nested layouts that require manual refactoring to be fully resolved.
\end{itemize}

\begin{table}[!ht]
\centering
\caption{Detailed remediation results for Dataset A (12 websites)}
\label{tab:individual_results}
\scalebox{0.85}{
\begin{tabular}{crrr}
\toprule
\textbf{ID} & \textbf{Initial Errors} & \textbf{Fixed Errors} & \textbf{RR (\%)} \\
\midrule
1  & 26  & 26  & 100.00\% \\
2  & 27  & 27  & 100.00\% \\
3  & 28  & 28  & 100.00\% \\
4  & 58  & 58  & 100.00\% \\
5  & 54  & 54  & 100.00\% \\
6  & 50  & 50  & 100.00\% \\
7  & 68  & 32  & 47.06\%  \\
8  & 21  & 10  & 47.62\%  \\
9  & 16  & 16  & 100.00\% \\
10 & 42  & 42  & 100.00\% \\
11 & 21  & 5   & 23.81\%  \\
12 & 10  & 4   & 45.71\%  \\
\midrule
\textbf{Average} & \textbf{35} & \textbf{29} & \textbf{80.35\%} \\
\bottomrule
\end{tabular}
}
\end{table}

\subsection{RQ2: Angular SPAs Remediation Efficacy}
In this case we used our tool for Dataset B (the 6 open-source Angular projects in Table~\ref{tab:dataset_b_projects}), whose results are shown in Table~\ref{tab:angular_results}.
One of the  goals in this experiment is measuring Build Integrity. After completing the remediation process, tests verified that all updated projects (100\%) compiled successfully using the command \verb|ng build --configuration production|. These findings demonstrate that the Abstract Syntax Tree-based injection technique maintains the syntactic validity of both TypeScript code. Our tool was able to obtain an average remediation rate of 86.0\% across all analyzed projects. These results, which range from 70\% to 100\%, highlight significant differences in each codebase's internal organization.
Projects dominated by static templates reached near-complete remediation, with virtually all detected issues corrected. In contrast, applications with more complex, highly dynamic behavior exhibited lower coverage, as the tool intentionally skipped ambiguous cases to avoid introducing regressions into the application's runtime behavior.

\begin{table}[!ht]
\centering
\caption{Remediation results for Angular open-source projects (Dataset B)}
\label{tab:angular_results}
\scalebox{0.85}{ 
\begin{tabular}{crrr}
\toprule
\textbf{ID} & \textbf{Initial Errors} & \textbf{Fixed Errors} & \textbf{RR (\%)} \\
\midrule
1 & 14  & 14  & 100.00\% \\
2 & 32  & 29  & 90.63\%  \\
3 & 66  & 46  & 69.70\%  \\
4 & 7   & 7   & 100.00\% \\
5 & 3   & 3   & 100.00\% \\
6 & 48  & 47  & 97.92\%  \\
\midrule
\textbf{Average} & \textbf{28.33} & \textbf{24.33} & \textbf{86.04\%} \\
\bottomrule
\end{tabular}
}
\end{table}

\subsection{RQ3: Semantic Preservation}
To address RQ3, we conducted a Semantic Functional Verification (SFV) to determine if the automated remediation introduced any destructive fixes. Following the side-by-side comparison methodology, we audited the visual and functional state of each project post-remediation.

In Dataset A, the visual structure remained intact in the majority of cases. We observed that the tool successfully injected ARIA attributes and updated contrast ratios without disrupting the CSS grid or flexbox layouts. However, on two websites (IDs 8 and 11), we identified minor visual regressions: adding descriptive labels slightly expanded the container dimensions, though the core visual hierarchy was preserved across the 12 analyzed sites.

In Dataset B, the results confirmed that the AST-based injection effectively shielded the application's logic across the 6 selected projects. Functional testing of the Angular projects showed a 100\% retention of interactive behavior. Even in projects with lower remediation rates, such as angular-realworld (ID 5), the tool's conservative stance proved successful; it prioritized maintaining data-binding and event listener integrity over risky HTML modifications. Consequently, no instances of broken navigation or disabled form elements were recorded, suggesting that the framework successfully balances accessibility compliance with the need to maintain the original UI's stability and design.

A cross-analysis of both datasets reveals that the tool's performance is strictly correlated with the \textbf{semantic integrity of the original codebase}. In Dataset A, the efficacy dropped significantly in legacy sites relying on nested tables. Similarly, in Dataset B, the tool adopted a conservative stance when facing complex, state-dependent variable bindings (Case 5).

\section{Conclusions and Future Work}
\label{sec:conclusions}

The experimental validation across 18 different scenarios (12 public websites and 6 open-source Angular projects) demonstrates the robustness of the approach. The system achieved a consistent remediation efficacy, averaging {80\% in static web pages (Dataset A)} and {86\% in Angular single-page applications (Dataset~B)}. Beyond remediation efficacy, we analyzed the computational effort required by the tool to perform these corrections. For the public websites in Dataset A, the average execution time was 15 minutes and 45 seconds per site. This duration includes the full pipeline of detection, multimodal image analysis via GPT-4o, and DOM stabilization. In contrast, remediation of Angular projects in Dataset~B required an average of 17 minutes and 3 seconds. The execution times for the latter were noticeably more consistent across projects, despite having a somewhat higher mean. This runtime performance suggests that the tool is quite feasible for incorporation into automated pipelines (e.g., in nightly builds) or as an autonomous agent.

The results in this paper validate two key hypotheses. On one hand, our tool is effective for rapid, non-invasive correction of static web pages, demonstrating high efficacy across modern HTML implementations despite challenges posed by legacy structures. On the other hand, the tool is also able to automate accessibility refactoring of SPAs without compromising software stability. The achievement of 100\% Build Integrity across all modified Angular projects is a critical milestone, suggesting that LLM-driven remediation is mature enough for inclusion in Continuous Integration (CI) pipelines. 


To further bridge the gap between automated suggestions and developer workflows, future research should focus on extending the static analysis engine to other ecosystems. Specifically, the immediate roadmap includes the implementation of support for React, adapting the AST injection strategy to handle JSX syntax and hook-based state management.

\begin{credits}
\subsubsection*{Declaration of Generative AI in the Scientific Writing Process}
During the preparation of this work, the authors used Large Language Models (LLMs) to enhance the technical and structural quality of the manuscript. Specifically, AI tools were used for language correction and translation, ensuring that technical descriptions and academic tone met international publication standards. Furthermore, these models assisted in calculating the Remediation Rate (RR) averages for both datasets, ensuring mathematical consistency throughout the results section. Finally, artificial intelligence (AI) was used to structure the Research Questions (RQs) and refine the overall hierarchy and flow of the article. The authors assume full responsibility for the final content and integrity of the research presented.
    
\end{credits}

\appendix
\section*{Appendix}
\label{sec:appendix_prompts}
This appendix provides the full text of the specialized prompts used in our tool.

\FloatBarrier
\begin{figure}[htbp]
\centering
    \begin{minipage}{0.8\linewidth} 
        \begin{promptbox}[Abstract Prompt Structure]{lightblue}
        \small 
        \textbf{[SYSTEM INSTRUCTION]} \\
        Act as a web developer expert in accessibility (WCAG 2.2 Level A+AA)... \\
        \textbf{[CONTEXT INJECTION]} \\
        - VIOLATION: \{\{axe\_description\}\} \\
        - SOURCE CODE: \{\{html\_or\_angular\_snippet\}\} \\
        \textbf{[OUTPUT CONSTRAINTS]} \\
        Return ONLY the fixed code wrapped in \verb|<<<TEMPLATE>>>| delimiters.
        \end{promptbox}
    \end{minipage}
\caption{Remediation prompt template used in our reparation tool.}
\label{lst:prompt_structure}
\end{figure}

\begin{figure}[htbp]
\centering
\begin{minipage}{0.85\linewidth}
    \begin{promptbox}[Multimodal Image Description (GPT-4o)]{lightblue}
    Describe this image for a web page alternative text ('alt'). Be concise and helpful for a person with a visual impairment.
    \end{promptbox}
\end{minipage}
\caption{Prompt used for multimodal image description.}
\label{lst:prompt_vision}
\end{figure}

\begin{figure}[htbp]
\centering
\begin{minipage}{0.85\linewidth}
    \begin{promptbox}[Color Contrast Correction (static web pages)]{lightgreen}
    Fix THIS color contrast error in the following HTML fragment. \\
    \textbf{VIOLATION:} \{\{description\}\} \\
    \textbf{QUICK RULES:} 
    \begin{itemize}
        \item Adjust ONLY the text color.
        \item Maintain backgrounds and layout exactly.
    \end{itemize}
    \textbf{FRAGMENT TO FIX:} \{\{original\_fragment\}\}
    \end{promptbox}
\end{minipage}
\caption{Prompt used for color contrast correction in static web pages.}
\label{lst:prompt_contrast}
\end{figure}

\begin{figure}[htbp]
\centering
\begin{minipage}{0.85\linewidth}
    \begin{promptbox}[General Remediation (Static Web Pages)]{lightgreen}
    Fix THIS accessibility error. \\
    \textbf{RULES:} Add aria-labels, alt text, or fix focusable elements according to Axe help text: \{\{help\_text\}\}. \\
    Return ONLY the fixed HTML fragment.
    \end{promptbox}
\end{minipage}
\caption{Prompt used for fixing for all issues in static web pages except color contrast.}
\label{lst:prompt_general}
\end{figure}

\begin{figure}[htbp]
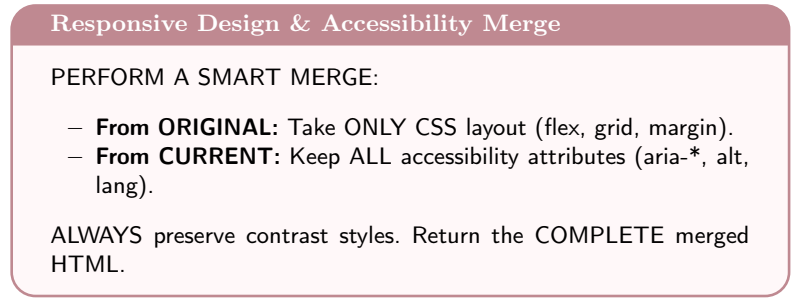

\centering
\begin{minipage}{0.85\linewidth}
    \begin{promptbox}[Responsive Design \& Accessibility Merge]{lightpink}
    PERFORM A SMART MERGE: 
    \begin{itemize}
        \item \textbf{From ORIGINAL:} Take ONLY CSS layout (flex, grid, margin).
        \item \textbf{From CURRENT:} Keep ALL accessibility attributes (aria-*, alt, lang).
    \end{itemize}
    ALWAYS preserve contrast styles. Return the COMPLETE merged HTML.
    \end{promptbox}
\end{minipage}
\caption{Prompt used responsive design and accessibility merge.}
\label{lst:prompt_responsive}
\end{figure}

\begin{figure}[htbp]
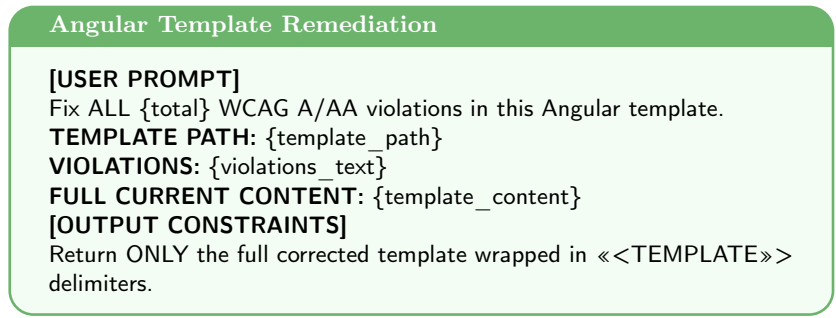

\centering
\begin{minipage}{0.9\linewidth}
    \begin{promptbox}[Angular Template Remediation]{lightgreen}
    \raggedright 
    \textbf{[USER PROMPT]} \\
    Fix ALL \{total\} WCAG A/AA violations in this Angular template. \\
    \textbf{TEMPLATE PATH:} \{template\_path\} \\
    \textbf{VIOLATIONS:} \{violations\_text\} \\
    \textbf{FULL CURRENT CONTENT:} \{template\_content\} \\
    \textbf{[OUTPUT CONSTRAINTS]} \\
    Return ONLY the full corrected template wrapped in <<<TEMPLATE>>> delimiters.
    \end{promptbox}
\end{minipage}
\caption{Prompt used for Angular SPAs remediation.}
\label{lst:prompt_template}
\end{figure}

\begin{figure}[htbp]
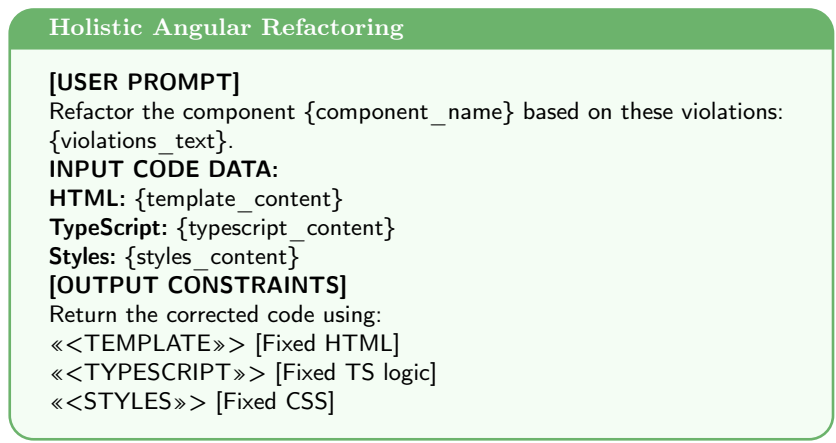

\centering
\begin{minipage}{0.9\linewidth}
    \begin{promptbox}[Holistic Angular Refactoring]{lightgreen}
    \raggedright 
    \textbf{[USER PROMPT]} \\
    Refactor the component \{component\_name\} based on these violations: \{violations\_text\}. \\
    \textbf{INPUT CODE DATA:} \\
    \textbf{HTML:} \{template\_content\} \\
    \textbf{TypeScript:} \{typescript\_content\} \\
    \textbf{Styles:} \{styles\_content\} \\
    \textbf{[OUTPUT CONSTRAINTS]} \\
    Return the corrected code using: \\
    <<<TEMPLATE>>> [Fixed HTML] \\
    <<<TYPESCRIPT>>> [Fixed TS logic] \\
    <<<STYLES>>> [Fixed CSS]
    \end{promptbox}
\end{minipage}
\caption{Prompt used for holistic Angular SPAs refactoring.}
\label{lst:prompt_angular}
\end{figure}

\FloatBarrier

%
%
%
\bibliographystyle{splncs04}
\bibliography{mybibliography}

\end{document}

%% file: config.tex
\usepackage{xcolor}
\usepackage[most]{tcolorbox}
\usepackage{graphicx}

\definecolor{lightblue}{RGB}{173, 216, 230}
\definecolor{lightpink}{RGB}{255, 182, 193}
\definecolor{lightgreen}{RGB}{144, 238, 144}

\newtcolorbox{promptbox}[2][]{%
    enhanced,
    colback=#2!10!white,
    colframe=#2!75!black,
    arc=3mm,
    title=#1,
    fonttitle=\bfseries, 
    fontupper=\sffamily\small, 
    boxrule=1pt,
    boxed title style={%
        colback=#2,
        colframe=#2!75!black,
        arc=1mm,
    }

\usepackage{cleveref}
\crefname{tcb@cnt@promptbox}{prompt}{prompts}
\Crefname{tcb@cnt@promptbox}{Prompt}{Prompts}
}